\documentclass[10pt]{article}
\usepackage[affil-sl]{authblk}
\usepackage{amsmath,amssymb,mathrsfs,graphicx}
\usepackage[english]{babel}
\title{Neutron reflectometry on highly absorbing films and its application to $\mathrm{^{10}B_4C}$-based neutron detectors}

\author[1,2,3]{F. Piscitelli \thanks{Corresponding author: francesco.piscitelli@esss.se}}
\author[1,2]{A. Khaplanov}
\author[4]{A. Devishvili}
\author[1,5]{S. Schmidt}
\author[1,5]{C. H\"{o}glund}
\author[5]{J. Birch}
\author[2,6]{A.J.C. Dennison}
\author[2]{P. Gutfreund}
\author[1,7]{R. Hall-Wilton}
\author[2]{P. Van Esch}

\affil[1]{European Spallation Source (ESS ERIC), P.O. Box 176, SE-221 00 Lund, Sweden.}
\affil[2]{Institut Laue-Langevin (ILL), 71, Avenue des Martyrs, 38042 Grenoble, France.}
\affil[3]{Department of Physics, University of Perugia, Piazza Universit\`a 1, 06123 Perugia, Italy.}
\affil[4]{Ruhr-Universit\"at Bochum, 44780 Bochum, Germany.}
\affil[5]{Thin Film Physics Division, Link\"{o}ping University, SE-581 83 Link\"{o}ping, Sweden.}
\affil[6]{Department of Physics \& Astronomy,  Uppsala University, BP 516, SE-751 20 Uppsala, Sweden.}
\affil[7]{Mid-Sweden University, SE-851 70 Sundsval, Sweden.}

\begin{document}
\date{\today}
\maketitle 
\thispagestyle{empty} 
                             
\begin{abstract}
Neutron reflectometry is a powerful tool used for studies of surfaces and interfaces. In general the absorption in the typical studied materials can be neglected and this technique is limited to the measurement of the reflectivity only. In the case of strongly absorbing nuclei the number of neutrons is not conserved and the absorption can be directly measured by using the neutron-induced fluorescence technique which exploits the prompt particle emission of absorbing isotopes. This technique is emerging from soft matter and biology where highly absorbing nuclei, generally in very small quantities, are used as a label for buried layers. 
\\Nowadays the importance of highly absorbing layers is rapidly increasing, partially because of their application in neutron detection; a field that has become more and more active also due to the $\mathrm{^3He}$-shortage. In this manuscript we extend the neutron-induced fluorescence technique to the study of thick layers of highly absorbing materials; in particular $\mathrm{^{10}B_4C}$. The theory of neutron reflectometry is a commonly studied topic, however the subtle relationship between the reflection and the absorption of neutrons is not widely known, in particular when a strong absorption is present. The theory for a general stack of absorbing layers has been developed and compared to measurements. This new technique has potential as a tool for characterization of highly absorbing layers.
\\We also report on the requirements that a $\mathrm{^{10}B_4C}$ layer must fulfill in order to be employed as a converter in neutron detection. 
\end{abstract}
\bigskip
\textbf{\\Keywords}
\medskip
\\ \textit{neutron-induced fluorescence, neutron reflectometry, Boron-10, neutron detection.}
\bigskip
\section{Introduction}
Neutron reflectometry is a powerful tool used for studies of surface chemistry, surface magnetism and solid films~\cite{cubitt3}. The reflection of neutrons was first demonstrated by Fermi and Zinn in 1944~\cite{fermizinn}. In most cases the absorption in the media can be neglected and the total number of neutrons remains constant via the sum rule: 
\begin{equation}
1-R-T = 0
\end{equation}
where R and T are reflectivity and transmittance normalised to the incident number of neutrons. Therefore most interfacial investigations are limited to the measurement of the reflectivity only. Measuring the transmittance in some cases can provide extra information about absorption or other anomalous scattering in the sample. However, the measurement of transmittance is often complicated due to refraction of transmitted neutrons and secondary scattering/absorption in the sample holder. 
\\The absorption on the other hand can be in some cases estimated by measuring the prompt $\alpha$ or $\gamma$ particle response, i.e. the neutron-induced fluorescence. The following isotopes are suitable for such investigations: $\mathrm{^{3}He}$, $\mathrm{^{6}Li}$, $\mathrm{^{10}B}$, $\mathrm{^{149}Sm}$, $\mathrm{^{151}Eu}$, $\mathrm{^{156}Hg}$, $\mathrm{^{155}Gd}$, $\mathrm{^{157}Gd}$. Several studies have been realised through ($\mathrm{n},\alpha$) reaction on $\mathrm{^{6}Li}$~\cite{padlo} or ($\mathrm{n},\gamma$) reaction on $\mathrm{^{155}Gd}$ and $\mathrm{^{157}Gd}$~\cite{nistbr}. The primary interest in this technique emerges from the fields of soft condensed matter physics and biology where the use of labelled molecules can allow enhanced sensitivity to adsorption of low concentrations while also providing the structural information normally associated with neutron reflectometry~\cite{manu2}. In these studies only a small fraction of the sample is composed of absorbing nuclei and the information on the buried labeled molecules is given by the neutron-induced fluorescence below the critical edge. 
\\In this manuscript we extend the neutron-induced fluorescence technique to the study of dense layers of highly absorbing materials; in particular on $\mathrm{^{10}B_4C}$ layers. Specifically, the number density of the absorbing media can be determined by measuring reflectivity and absorption simultaneously. The two sets of data have been fitted using a set of equations we derived from the theory we developed for a generic stack of a highly absorbing layers. The theory of neutron reflectometry is a commonly studied topic, however the subtle relationship between the reflection and the absorption of neutrons is not widely known, in particular when a strong absorption is present.
\\Nowadays the importance of $\mathrm{^{10}B_4C}$ layers is increasing. Most of the neutron sources in the world, such as the European Spallation Source (ESS)~\cite{esstdr,kir4} in Sweden, are necessarily pushing the development of their detector technologies, due to the increased flux available for the neutron scattering science and the scarcity of $\mathrm{^3He}$, the so-called ``Helium-3 crisis''~\cite{ChoScience,KarlNN}. $\mathrm{^{10}B}$ along with $\mathrm{^{3}He}$ and $\mathrm{^{6}Li}$ isotopes are the main actors in thermal neutron detection due to their large absorption cross-sections~\cite{blue}. Concerning small-area detectors ($\mathrm{<1m^2}$), the current detector technology is reaching fundamental limits in position resolution and rate capability. The Multi-Blade~\cite{buff3,framb,fratesi,mbmg}, the Jalousie detector~\cite{kleinjalousie,kleinjalousie2}, the A1CLD~\cite{nowak} and many others~\cite{vanvuure,gor} are an example of the detector developments for small area coverage which exploit solid neutron converters operated at grazing angle (between 0 and 10 deg) in order to increase the neutron detection efficiency. For detection applications the neutron reflection from the detector elements must be avoided because it limits the maximum efficiency that can be attained and it may give rise to misaddressed events.
\\Recently, high quality, low cost production of square meters of $\mathrm{^{10}B_4C}$~\cite{carina,nowak} become possible and some of the detector developments are focused on the application of such films in inclined geometry. 
\\Neutron reflectometry with neutron-induced fluorescence is a powerful tool to investigate the performance of highly absorbing layers employed in neutron detection. In this manuscript we also report on the requirements that a converter layer must fulfill to be employed in a detector to avoid reflection. We characterized the $\mathrm{^{10}B_4C}$ layer when deposited on both $\mathrm{Si}$ and $\mathrm{Al}$ substrates.
\subsection{Theory of neutron reflection on highly absorbing layers} 
\begin{figure}[!ht]
\centering
\includegraphics[width=8cm,keepaspectratio]{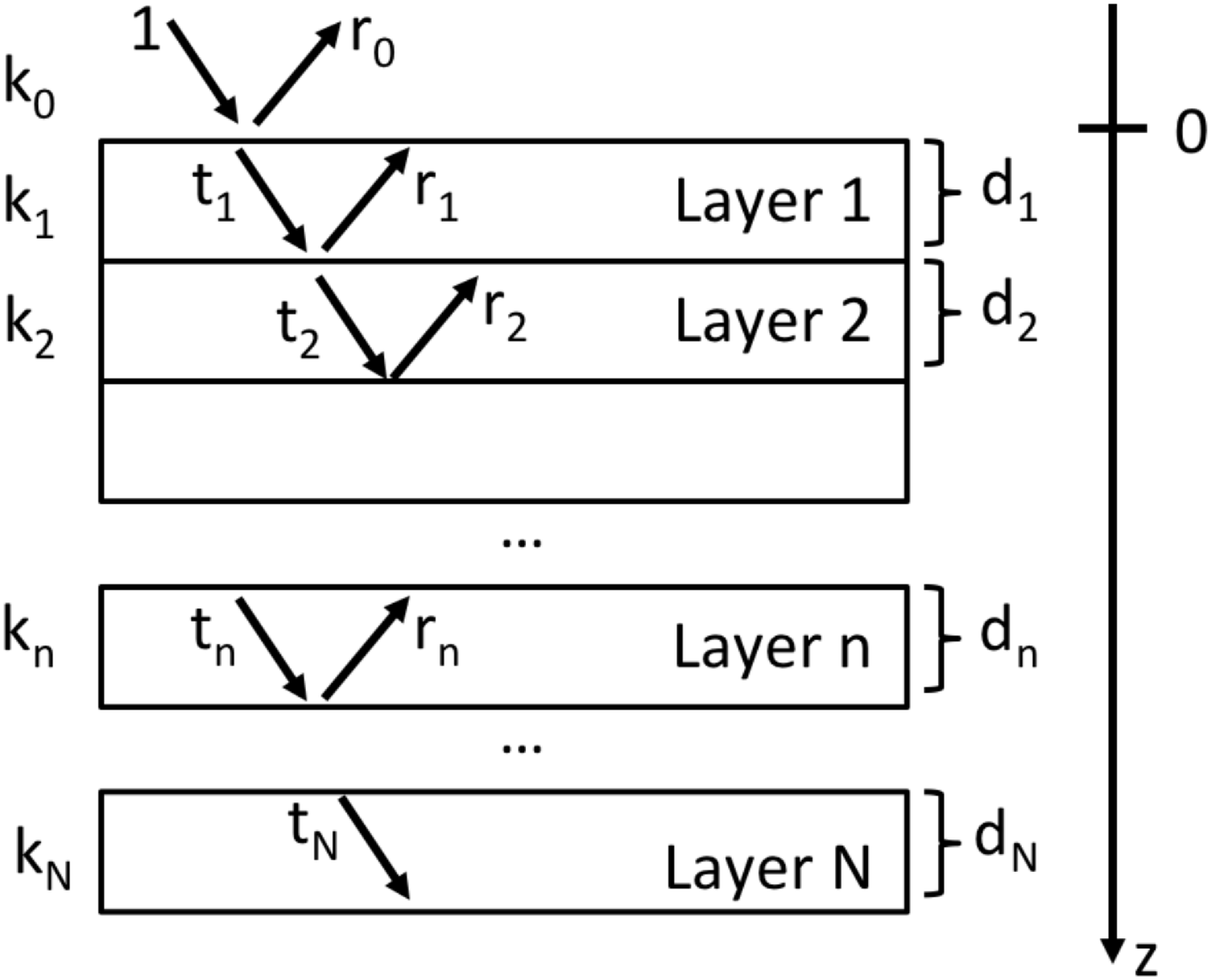}
\caption{\footnotesize A stack of N layers of thicknesses $d_n$; $k_n$ is the normal component of the wave-vector, $r_n$ and $t_n$ are the complex amplitudes of the wave-functions in the $n-th$ layer.}
\label{mlayer1}
\end{figure}

\begin{figure}[!ht]
\centering
\includegraphics[width=7.8cm,keepaspectratio]{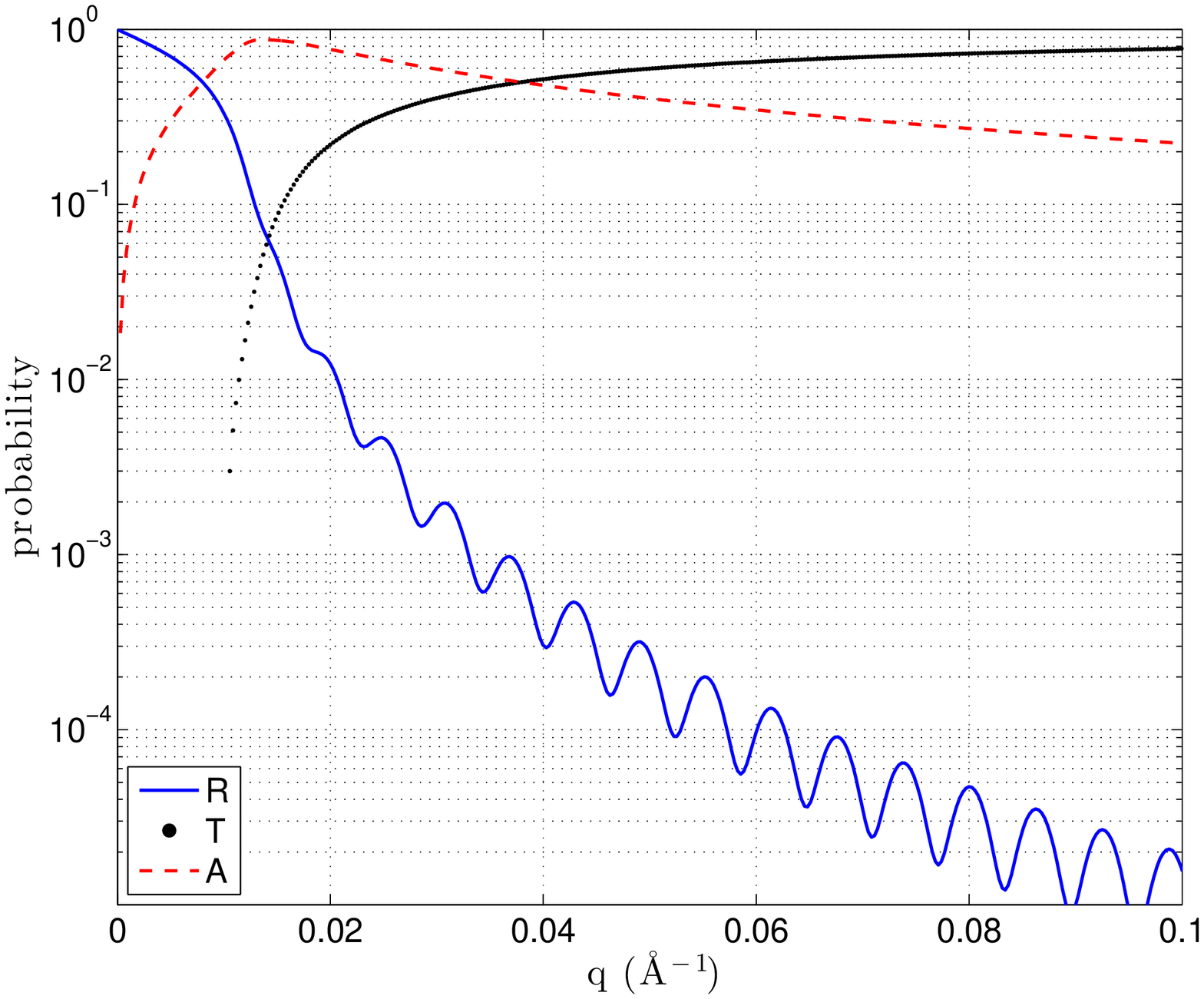}
\includegraphics[width=7.8cm,keepaspectratio]{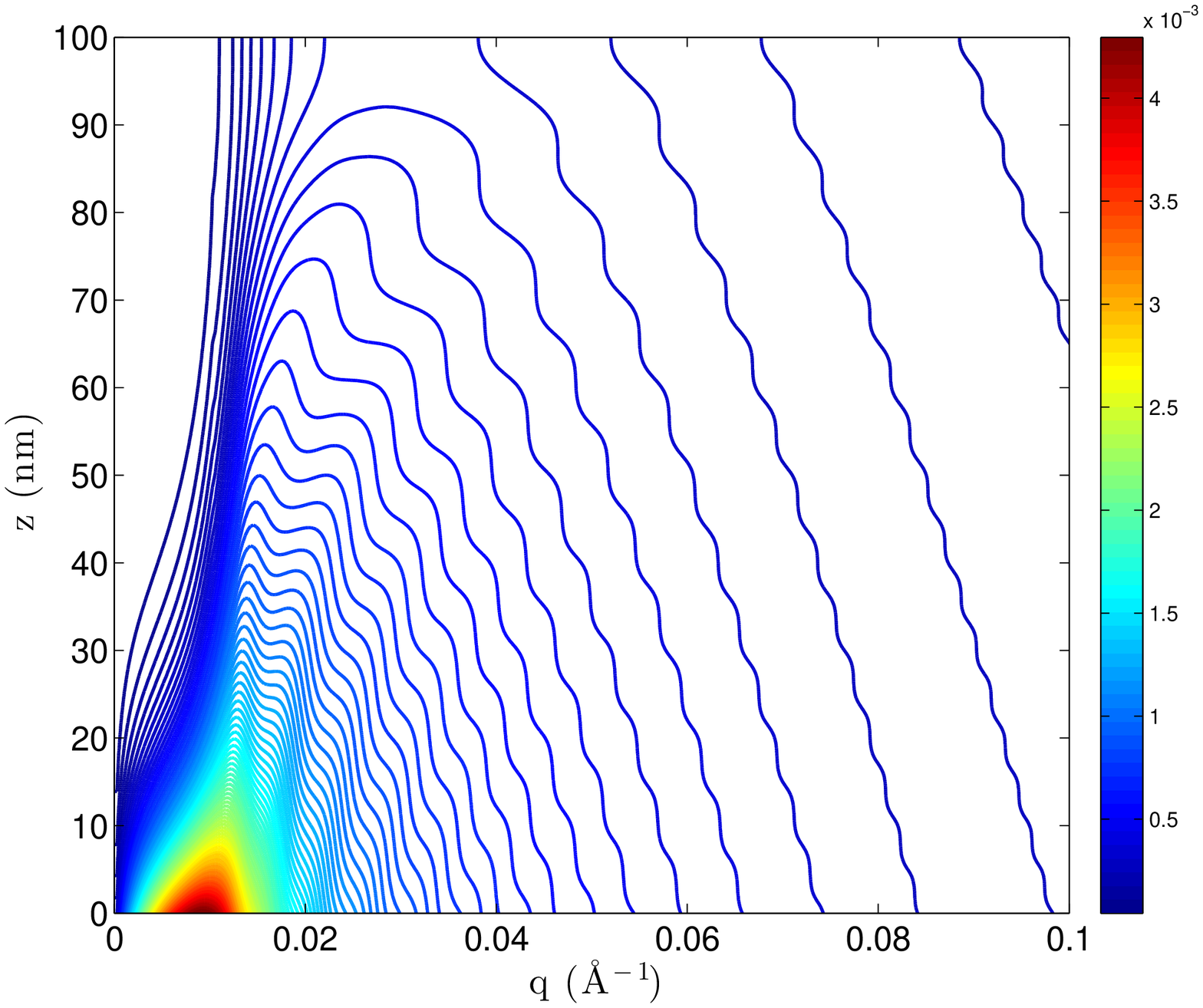}
\caption{\footnotesize The calculated Reflectivity (R), Transmission (T) and Absorption (A) for a $100 \, nm$ $\mathrm{^{10}B_4C}$ layer deposited on Si (left), the probability for a neutron to be absorbed in the layer as a function of $\mathrm{z}$ and $\mathrm{q}$, i.e. the quantity $- \frac{4 \pi}{k_0} |Y_z|^2 N_b^{im}$ (right); the color scale is expressed in \AA$^{-1}$.}
\label{borplo1}
\end{figure}

The reflection of neutrons from surfaces is a phenomenon caused by the change of refractive index across the interface analogous to that for light. The analysis of specular neutron reflectivity reveals the nuclear density profile perpendicular to the surfaces and interfaces. The sensitivity of neutron reflectivity to interfaces is due to the fact that the kinetic energy of a neutron projected on the normal to the surface at grazing incidence is comparable with the potential energy of the reflecting interface, $V$. At the same time, the wavelength corresponding to this component of the kinetic energy matches often the thickness of thin films of interest well.
\\ When dealing with neutron absorbers, the theory describing the physical process of reflection has to be modified in order to, not only take the possibility for a neutron to be scattered into account, but also its absorption by nuclei. The sum rule in this case will take the following form: 
\begin{equation}\label{sumr}
1-R-T-A = 0
\end{equation}
where R, T and A are respectively reflectivity, transmittance and absorption normalised to the incident number of neutrons. The scattering length of a nucleus is a complex quantity. Its real and imaginary parts can be associated to the scattering process but only its imaginary part to the absorption~\cite{sears,hayter}.
\\ The scalar potential $V$ in the Schr\"{o}dinger equation, $-\frac{\hbar^2}{2 m_n}\nabla^2 \Psi + V \Psi = E \Psi$, will contain the contribution given by the absorption, and it is:
\begin{equation}
V=\frac{2\pi\hbar^2}{m_n}\left(N_b^{real} + i \, N_b^{im}\right)
\end{equation}
where $m_n$ is the neutron mass and where the scattering length density $N_b$ can be calculated according to $N_b=\sum_i b_i n_i$ (with $b_i$ the scattering length of the $i-th$ species, isotope, and $n_i$ its number density). The solutions of the Schr\"{o}dinger equation, with a complex potential, can still be written as for a real potential~\cite{cubitt3}, however the wave-vectors will be complex quantities. The solutions of the Schr\"{o}dinger equation can be still factorized as two plane waves, one orthogonal to the surface and one parallel. The complex potential the neutrons experience affects only the normal component of the momentum. Continuity of wave-functions and their derivatives at boundaries lead to the unaltered conservation of the parallel wave-vector. Thus, we refer only to the normal component of the wave-functions and we denote the wave-vector in a generic layer $n$ simply as $k_{n\bot}=k_n$.
\\The reflectivity profile can be measured in Time-Of-Flight (ToF) or in monochromatic mode without affecting the results (if we can assume constant imaginary scattering length density, meaning there are no absorption resonances). By denoting with $k_0$ the normal component of the incoming wave-vector, the reflectivity only depends on $\mathrm{\theta}$ and $\mathrm{\lambda}$ through $k_0=\frac q 2=\left(2\pi/\mathrm{\lambda}\right)\sin(\mathrm{\theta})$. Thus any method (ToF or monochromatic) is used to get a value for the measured reflectivity the result is the same for same $\mathrm{q}$. \\The change in the normal wave-vector has an imaginary part given by the complex potential that results into an exponentially reduced amplitude of the wave-function~\cite{hayter}. As such, the absorption has been taken into account in the amplitudes of the wave-functions. An equivalent way of looking at the absorption is as a negative source term that appears in the continuity equation. This gives an indication of where the absorption, which is included in the wave-functions, takes place. If there would not have been any imaginary potential, probability is conserved in quantum mechanics, and there are no source terms in the continuity equation. The source term is then entirely related to the imaginary part of the potential. The continuity equation gives the normalization of the wave-functions; in the region where the material is an absorber it has to be generalized considering the probability for a neutron to be absorbed~\cite{schiff}. The continuity equation can be written:
\begin{equation}\label{eq4re}
\frac{\partial P(\bar{r},t)}{\partial t}+\nabla \cdot J(\bar{r},t)= - \frac{2}{\hbar} P(\bar{r},t) Im\left\{V\right\} \qquad \Longrightarrow  \qquad \nabla \cdot J(\bar{r},t)= - \frac{4 \pi \hbar}{m_n} P(\bar{r},t) N_b^{im}
\end{equation}
where $P(\bar{r},t)$ and $J(\bar{r},t)$ are the probability density function and the probability current respectively; assuming stationary conditions: $\frac{\partial P(\bar{r},t)}{\partial t}=0$. 
\\We can use Equation~\ref{eq4re} to determine the amount of absorption in a certain volume, for instance, the entire layer of absorbing material with thickness $d$.  Mathematical consistency between Equation~\ref{sumr} and~\ref{eq4re} then requires that the following relation holds:
\begin{equation}\label{eq4}
A = \frac{m_n}{\hbar k_0}\int_0^d \nabla \cdot J(z,t) \, dz = -\frac{4 \pi}{k_0} \int_0^d P(z,t) N_b^{im} \, dz = - \frac{4\pi}{k_0} \int_0^d \left| Y_z\right|^2 N_b^{im} \,dz
\end{equation}
where $k_0$ the normal component of the incoming wave-vector, $d$ is the thickness of the absorbing layer, $N_b^{im}$ is the imaginary part of the scattering length density of the absorbing medium and $Y_z$ is the solution of the Schr\"{o}dinger equation in the absorbing layer. The solutions of the Schr\"{o}dinger equation in the different regions can be calculated iteratively according to the Parratt formalism~\cite{parratt}. 
\\In order to generalize to a multi-layer system, let us consider a stack of N layers of thicknesses $d_n$; $k_n$ is the normal component of the wave-vector in each region, $r_n$ and $t_n$ are the complex amplitudes of the wave-functions in the $n-th$ layer (see Figure~\ref{mlayer1}). In the first medium (generally air) the incoming amplitude is 1 and the reflected amplitude is $r_0$. We consider the amplitude in the layer $N$ to be only transmitted since we assume it to be a substrate of infinite thickness. In the generic region $n$ the normal component of the solution of the Schr\"{o}dinger equation is: 
\begin{equation}
Y_n(z)=t_n\,e^{+i\,k_n\,z}+r_n\,e^{-i\,k_n\,z}
\end{equation}
Note that not all the layers must necessarily be an absorber. With absorber we mean here a material which has an absorption cross-section that can not be neglected with respect to the scattering cross-section ($\sigma_a\sim\sigma_s$). For most materials the absorption cross-section is a few orders of magnitudes smaller than the scattering cross-section. The total absorption is given by the sum of the contributions of the single layers. Equation~\ref{eq4} can be used to calculate the absorption of each layer considering the solution $Y_n$ in the region of thickness $d_n$ and the respective imaginary part of the scattering length density $N_b^{im}$.
\\ Let us consider a finite thickness as in~\cite{nistbr}, $d$, of absorbing material deposited on a substrate, e.g. Si. In this specific case we consider $N=2$. We identify three regions delimited by two interfaces: air/absorber ($z=0$) and absorber/substrate ($z=d$). We denote $k_{n\bot}=k_n$ with $n=0,1,2$, the normal component of the wave-vectors in the three regions is defined by the potentials $V_n$.
\\ The reflection and transmission probabilities are given by:
\begin{equation}
R = \frac{\left|J_r\right|}{\left|J_i\right|}, \qquad  \qquad T =
\frac{\left|J_t\right|}{\left|J_i\right|}
\end{equation}
with
\begin{equation}
\left|J_i\right|=\frac{\hbar\,k_0}{m_n}, \qquad \left|J_r\right|=\frac{\hbar\,k_0}{m_n}\left(r_0 \cdot r_0^\ast\right), \qquad \left|J_t\right|=\frac{\hbar\,k_2}{m_n}\left(t_2 \cdot t_2^\ast\right)
\end{equation}
where $J_i$,  $J_r$ and $J_t$ are the probability current of the incoming, reflected and transmitted waves respectively. 
The measured reflectivity, the transmission inside the substrate and the absorption in the layer are:
\begin{equation}\label{eqaf22}
\begin{aligned}
R&=r_0 \cdot r_0^\ast\\
T&=\frac{k_2}{k_0}\left(t_2 \cdot t_2^\ast\right)\\
A&=1-R-T=\frac{1}{k_0}\int_0^d \nabla \cdot J_1(z,t) \, dz = - \frac{4 \pi}{k_0} \int_0^d |Y_z|^2 N_b^{im} \, dz
\end{aligned}
\end{equation}
where $J_1$ is the probability current calculated for $Y_z$, and $r_0$, $t_2$ are the amplitudes of the waves in the first and third medium. The three different expressions in the third line of Equation~\ref{eqaf22} are three mathematically equivalent ways to calculate the absorption that results from the introduction of an imaginary part to the potential.
\\ As example we take a $\mathrm{^{10}B_4C}$ layer ($N_b = (2.5-1\cdot i)\cdot10^{-6}\,$\AA$^{-2}$) of $d=100 \, nm$ deposited on Si ($N_b = 2.14\cdot10^{-6}\,$\AA$^{-2}$). In Figure~\ref{borplo1}, on the left we show reflectivity, transmission and absorption, as calculated in Equation~\ref{eqaf22} as a function of $\mathrm{q}$. On the right we show the probability for a neutron, carrying a given $\mathrm{q}$, to be absorbed at certain depth in the layer, i.e. the quantity $- \frac{4 \pi}{k_0} |Y_z|^2 N_b^{im}$. We notice that absorption increases in proximity of the critical edge ($\mathrm{q_c}$) that is, in this case, at about $q_c =0.01\,$\AA$^{-1}$.
\section{Experimental setup}
\subsection{Preliminary characterization of the $\mathrm{^{10}B_4C}$-layers}
$\mathrm{^{10}B_4C}$ thin films were deposited in Link\"{o}ping University by the Thin Film Physics Division in an industrial deposition chamber (CemeCon AG, Germany) using direct current magnetron sputtering (DCMS). The films were synthesized from enriched $\mathrm{^{10}B_4C}$ sputter targets in an Ar discharge. The sputter targets were enriched to a $\mathrm{^{10}B}$ content of about $98\%$ (specified by the supplier) of the total Boron content~\cite{carina,carinatesi,carrad}.
\\A thin layer of about $100\,\mathrm{nm}$ $\mathrm{^{10}B_4C}$ film was deposited on Si(001). A thick layer of about $1\,\mu\mathrm{m}$ was deposited on Si(001) and aluminium (EN AW-5083) substrates, the latter is a common substrate used in neutron detectors~\cite{framb,mg}. 
\\The chemical composition of the film were investigated by X-ray photoelectron spectroscopy (Axis UltraDLD, Kratos Analytical, UK) using monochromatic X-ray radiation ($h_{\nu} = 1486.6\,\mathrm{eV}$). The base pressure in the analysis chamber during acquisition was $<10^{-7}\,\mathrm{Pa}$. XPS is a surface sensitive technique; data contains information of the material surface and sub-surface layers of approximately $10\,\mathrm{nm}$ in depth. The XPS core level spectra of the $B1s$, $Ar2p$, $C1s$ and $O1s$ regions were recorded on the as-received sample and after each $\mathrm{Ar^{+}}$ etching step during depth profiling. Depth profiling was performed with an $\mathrm{Ar^{+}}$ ion beam, rastered over an area of $3\times3\,\mathrm{mm^2}$ at an incidence angle of $20^{\circ}$. The $\mathrm{Ar^{+}}$ etch sequence removed $5\,\mathrm{nm}$ of material in each of the first two steps, corresponding to a sputter etch time of $85\,\mathrm{s}$ per step. The following steps removed each $10\,\mathrm{nm}$ of $\mathrm{^{10}B_4C}$ corresponding to a sputter etch time of $171\,\mathrm{s}$. The composition of the $100\,\mathrm{nm}$ thin film was extracted after each $\mathrm{Ar^{+}}$ etch step. Here, the corresponding core level spectra were evaluated after subtracting a Shirley-type background using the CasaXPS software and elemental cross sections provided by Kratos Analytical. After removing the first 10 nm of oxidized film surface, the composition in the bulk of the film did not vary by more than $1\,\%$. Table~\ref{figxps1} shows the composition of the $\mathrm{^{10}B_4C}$ layer as a function of the sampling depth. It shows that there is a thin ($\sim5\,\mathrm{nm}$) partially oxidized layer on the surface of the $\mathrm{^{10}B_4C}$ layer.
\begin{table}[!ht]
\caption{\footnotesize $\mathrm{^{10}B_4C}$ film composition with progressing film depth from the surface as obtained from relevant XPS core level spectra. The error on the composition never exceeeds $\pm 2\%$.}\label{figxps1}
\centering
\begin{tabular}{|c|c|c|c|c|}
\hline \hline
depth ($\mathrm{nm}$) & $\mathrm{^{10}B}\,(at.\%)$ & $\mathrm{^{11}B}\,(at.\%)$ & $\mathrm{C}\,(at.\%)$ & $\mathrm{O}\,(at.\%)$ \\
\hline
$0$        & $38.2$ & $1.5$ & $41.5$ & $18.8$ \\
$5$        & $75.8$ & $1.5$ & $19.6$ & $3.1$ \\
$\geq 10$      & $77.3$ & $1.5$ & $19.3$ & $1.9$ \\
\hline \hline
\end{tabular}
\end{table}
In order to accesses the surface roughness of the $\mathrm{^{10}B_4C}$ film, Atomic Force Microscope (AFM) scans were performed in tapping mode using a Digital instruments Multimode equipped with a silicon tip (type: PPP-NCHR, Nanosensors, Swizerland) with a resonance frequency of 284 kHz. Four measurements with a scan size $1\,\mathrm{\mu m} \times 1\,\mathrm{\mu m}$ on different sample locations were performed. The roughness of the of the $100\,\mathrm{nm}$ thick $\mathrm{^{10}B_4C}$ film on Si when excluding surface particles was extracted to be $(0.47\pm0.02)\mathrm{nm}$. The roughness when including the surface particles was $(6.12\pm0.02)\mathrm{nm}$.
\\ X-Ray Reflectivity (XRR) was performed to determine the layer density using a Philips X’Pert Pro MRD diffractometer equipped with a hybrid mirror monochromator and a 2-bounce Ge 220 triple-axis crystal analyzer. The film density was determined to be $(2.45\pm 0.05)\,\mathrm{g/cm^3}$ by fitting the XRR data using a two layer model with the X’pert reflectivity software.
\\For the analysis of the reflectometry and fluorescence data we assume the amount of $\mathrm{^{11}B}$ fixed at $1.5\%$ . This assumption is made to reduce the number of free parameters. Moreover, $\mathrm{^{11}B}$ and $\mathrm{C}$ are not distinguishable in neutron reflectometry due to their similar scattering lengths~\cite{sears}, hence any percentage we fix of $\mathrm{^{11}B}$ can be considered as a part of the total $\mathrm{C}$ amount.
\\According to the composition (Table~\ref{figxps1}) and the mass density, we expect the scattering length density $N_b=\sum_i b_i n_i$ of the layers to vary in the range given in Table~\ref{tabnb}.
\begin{table}[!ht]
\caption{\footnotesize Scattering length density ($N_b$) of the sputtered $\mathrm{^{10}B_4C}$ layer calculated according to the depth and the composition given in Table~\ref{figxps1}. The minimum and the maximum values are calculated according to the range of the layer mass density.}\label{tabnb}
\centering
\begin{tabular}{|c|c|c|}
\hline \hline
depth ($nm$) & min $N_b$($10^{-6}$\AA$^{-2}$) &  max $N_b$($10^{-6}$\AA$^{-2}$)  \\
\hline
$0$    & $4.72-0.49\,i$  & $4.92-0.51\,i$ \\
$5$    & $2.06-1.10\,i$  & $2.14-1.14\,i$ \\
$\geq 10$  & $1.95-1.13\,i$  & $2.03-1.17\,i$ \\
\hline \hline
\end{tabular}
\end{table}
\\Note that all the elements in the film, including $\mathrm{^{10}B}$, contribute to the real part of $N_b$ and almost only the $\mathrm{^{10}B}$ amount determines its imaginary part. In fact the imaginary part of the $\mathrm{^{10}B}$ scattering length is about six order of magnitudes larger of that of any of the other components~\cite{sears}.
\subsection{Reflectivity and absorption measurements}
\begin{figure}[!ht] \centering
\includegraphics[width=8cm,keepaspectratio]{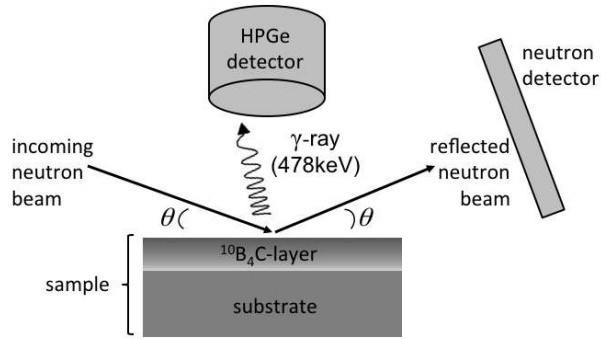}
\caption{\footnotesize Schematic of the experiment on SuperAdam. A HPGe-detector was used to measure the neutron absorption in the $\mathrm{^{10}B_4C}$ layer exploiting the capture of the the prompt $\gamma$-ray ($478\,\mathrm{keV}$) of the $\mathrm{n} + \mathrm{^{10}B}$ reaction.}
\label{figexpsupadam9046}
\end{figure}

Two experiments have been performed. The first set of data was taken using D17~\cite{cubittD17} at ILL which was used as a ToF reflectometer to preliminary quantify the actual reflectivity of the coatings and to have a direct measurement of the reflection as a function of the neutron wavelength for different angles. A second experiment has been performed on SuperAdam~\cite{superadam} at ILL, which is a monochromatic reflectometer in a setup that also included a $\gamma$-ray-spectrometer. The two experiments allow the comparison of the two techniques (ToF and monochromatic), and additionally provide information on neutron converter reflectivity and layer composition.
\\The $1\,\mathrm{\mu m}$-thick $\mathrm{^{10}B_4C}$ layers deposited on both Si and Al were measured on D17. Both $1\,\mathrm{\mu m}$ and $100\,\mathrm{nm}$ layers deposited on both Si and Al were measured on SuperAdam. 
\\On the D17 instrument, the reflectivity profiles were measured using three angles $\mathrm{\theta}=0.5^{\circ},1^{\circ},2^{\circ}$ in ToF-mode between $\lambda=2$\AA \, and $\lambda=25$\AA. The reflected intensity (and the direct beam) in ToF can be measured in energy dispersive mode by acquiring the full neutron wavelength spectrum at once. The reflectivity is calculated as the ratio between the reflected and the direct wavelength spectra. The background, uncorrelated with the instrument timing, was evaluated by looking at a region of the detector far from the specular reflection. This background has been subtracted from the reflected and the direct beam spectra. 
We repeated the neutron reflectivity measurement on SuperAdam in angle dispersive mode at fixed wavelength to $\lambda=4.4$\AA. The sample angle has been changed in order to measure reflectivity for the corresponding q. 
\\In addition to neutron reflectivity (R), the absorption (A) has been measured at the same time. A liquid nitrogen cooled HPGe-detector has been used to capture the prompt $\gamma$-ray response of $\mathrm{n} + \mathrm{^{10}B}$ reaction:
\begin{equation}
\begin{aligned}
 \mathrm{n} + \, \mathrm{^{10}B} \rightarrow \mathrm{^7Li^*} + \alpha  &\rightarrow \mathrm{^7Li} + \alpha +  \gamma \, (478\,\mathrm{keV}) &+  2.31 \,MeV  &\,\,(94\%)\\  
                                                             &\rightarrow  \mathrm{^7Li} + \alpha                           &+ 2.79 \,MeV   &\,\,(6\%)
\end{aligned}
\end{equation}
In the $94\%$ branch, a prompt $478 KeV$ $\gamma$-ray is emitted. The schematic representation of the experimental setup at SuperAdam is depicted in Figure~\ref{figexpsupadam9046}. Taking into account both the HPGe-detector efficiency and the solid angle, we estimate the absolute efficiency for the $478\,\mathrm{keV}$ $\gamma$-ray photo-peak detection to be about $5\%$. The HPGe-detector has been energy calibrated using a $\mathrm{^{22}Na}$ source.
For a given sample we record for each angle a spectrum for the HPGe-detector and a neutron detector image of the reflected neutrons. The normalization is given by the direct beam which was also recorded. 
\\Each point of the absorption curve (A) has been obtained by fitting the HPGe-detector spectrum, around the $478\,\mathrm{keV}$ $\gamma$-ray photo-peak, with a model that includes a linear background. The latter is subtracted from the actual number of counts and it takes the Compton background due to other $\gamma$-ray energies into accounts.
\\Since the sample length is not infinite there will be a certain point in the angular scan when the size of the beam coincides with the projected size of the sample, this is the so-called over-illumination angle ($\mathrm{\theta_{over}}$). Therefore the raw intensity of the reflection rises until $\mathrm{\theta_{over}}$ and then behaves as an absolute reflectivity profile. Hence a data correction was applied in order to transform the intensity of the reflection into reflectivity. 
\\The model considered to fit the data has been explained previously, the absorption and the reflected intensities are fitted simultaneously with a least square minimization. The scattering length densities (real and imaginary parts), layer roughnesses $\sigma_r$, layer thickness and HPGe-detector efficiency, are the free fitting parameters. The HPGe-detector efficiency takes the efficiency of the detector for the $478\,\mathrm{keV}$ $\gamma$-ray photo-peak and the solid angle subtended into account. 
\section{Results and discussion}
\begin{figure}[!ht]
\centering
\includegraphics[width=16cm,keepaspectratio]{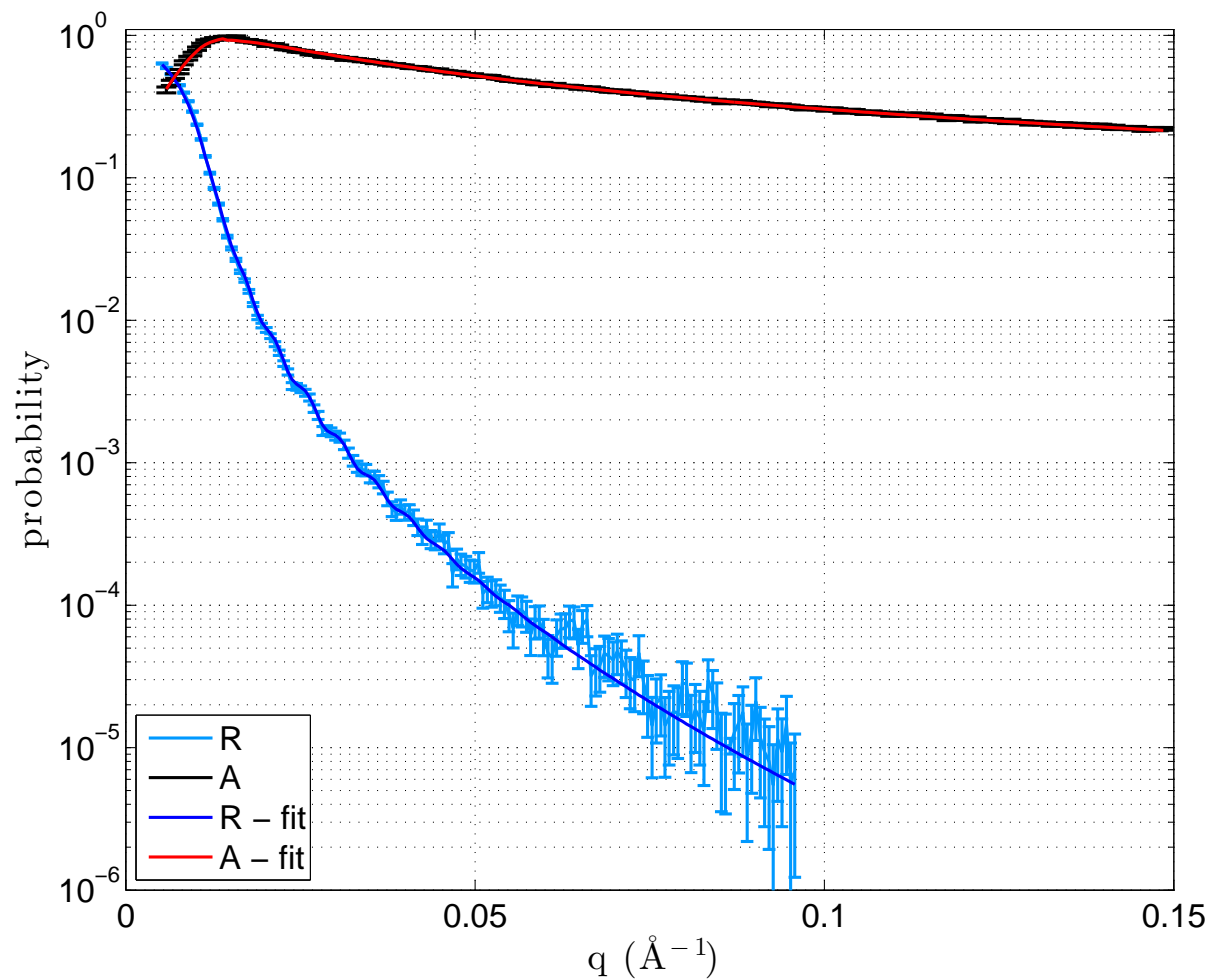}
\caption{\footnotesize Measured and fitted reflectivity (R) and absorption (A) probabilities for the $100\,\mathrm{n m}$ $\mathrm{^{10}B_4C}$ sample deposited on Si as a function of the momentum transfer (q).} \label{fig23}
\end{figure}

Figure \ref{fig23} shows the reflectivity and absorption profiles for the $100\,\mathrm{n m}$ $\mathrm{^{10}B_4C}$ sample deposited on Si. The over-illumination correction is here applied to the data to visualize the absolute reflectivity curve.
The sample of $1\,\mathrm{\mu m}$ deposited on Al has also been measured, but no specular reflection has been observed while absorption is comparable to that of the sample deposited on Si. The specular reflection is attenuated by the surface roughness of the film, which is for Al-substrates in the order of a few tens of nm. The Al substrate is etched and this to explain its large roughness. The high roughness is not a characteristic of Al in general, but it can also be very flat if manufactured in a different way.
\\Off-specular reflection was not observed in any sample (Si or Al) or it is below the background level. This suggests that there is no in-plane structure of the coating. By comparing this result to the AFM measurements we can conclude that the surface particles are randomly scattered over the surface without any correlation, as expected in a sputtering process. 
\\From Table~\ref{figxps1} we notice that a thin oxide layer is present on top of the converter layer. The scattering length densities vary significantly in a few nm (see Table~\ref{tabnb}). We can chose to fit the data of the $100\,\mathrm{n m}$ with a model that includes either two or three layers. In the first case we consider a unique converter layer deposited on a substrate and in the second case we also add an oxide layer at the air/converter interface. Since $1\,\mathrm{\mu m}$ is far beyond the sampling depth of any neutron reflectometer and any reflection from the substrate is attenuated by the $\mathrm{^{10}B_4C}$ layer, for the thicker sample we can exclude the substrate contribution in the model; we can fit the data with either one single layer or two. The adopted choice of model is important and has a significant influence on the interpretation of the result. The oxide layer can be modeled as a rough layer deeply overlapped with the $\mathrm{^{10}B_4C}$ layer below, since specular reflectivity cannot distinguish between a rough or a diffuse interface. We used a smearing function on the 1D scattering potential at each interface.
\\We fix the scattering length density of the Si-substrate to $2.14\cdot10^{-6}\,$\AA$^{-2}$, its roughness is a free fitting parameter. If we exclude the oxide layer the fit converges with a reduced chi-square of $\tilde{\chi}=1.7$, without setting any boundary condition on the parameters of the fit. In Table~\ref{tabd5680232} the fitting parameters are listed. We define the error on a parameter as the change of a parameter which produces a change of the reduced chi-square equal to 1. In case we include the oxidized layer, the fit may converge to many different local minima because of the number of free fitting parameters; it converges with $\tilde{\chi}=1.1$ only when the parameters are allowed to vary in physically acceptable boundaries based on the XPS measurements: e.g. the scattering length density is allowed to vary in the ranges given in Table~\ref{tabnb}. Both samples deposited on Si show similar results. 
\\Referring to Table~\ref{tabd5680232} in the case the oxide layer is excluded, the resulting $N_b$ is in the range of the value given in Table~\ref{tabnb}, but it does not represent the oxide nor the $\mathrm{^{10}B_4C}$ layer, rather it is the mixture of both. The result of the fit is physically correct only when the extra layer is included by the fit which is driven by complementary measurements. The reflectivity curve is produced by the interfaces and absorption curve is instead a volume effect; more precisely, from Equation~\ref{eqaf22} we can calculate that the absorption is mainly determined by a few hundreds of nm on the surface, while the oxidized layer plays a more significant role in the reflectivity curve. In the absorption curve the major contribution is represented by the $\mathrm{^{10}B_4C}$ underneath. Hence, if we do not include the oxide interface in the fit, the real part of the 1D potential increases. 
\\The absorption (A) is given by the imaginary part of the scattering potential and we can assume that it is entirely determined by the $\mathrm{^{10}B}$ content; thus from the imaginary part of $N_b$ the $\mathrm{^{10}B}$ number density is univocally determined. With $\operatorname{Im}(N_b)=1.1\cdot10^{-6}$\AA$^{-2}$, it is $n_{^{10}B}=0.103\,$\AA$^{-3}$. 
\\ The layer roughnesses are compatible to the result of the AFM measurement. The thinner layer thickness is estimated to be $(121\pm2)nm$. Since the oxidized layer is a diffuse interface into the $\mathrm{^{10}B_4C}$ we get from the fit a layer with a large relative roughness. 
\begin{table}[!ht]
\caption{\footnotesize Fitting parameters including results of the fit with and without the oxide layer on the sample surface. Values listed refer to $\mathrm{^{10}B_4C}$ deposited onto Si.} \label{tabd5680232}
\centering
\begin{tabular}{|c|c|llll|}
\hline \hline
sample             & model     & layer                      &  d(nm)     &  $N_b$($10^{-6}$\AA$^{-2}$) & $\sigma_r$(nm)  \\
\hline
$100\,\mathrm{nm}$ & 2 layers  &  1 ($\mathrm{^{10}B_4C}$)    &   $121\pm2$  & $(2.50\pm0.04)-(1.11\pm0.04)i$    & $4.2\pm0.8$  \\
                   &           &  2 (substrate)              &   -          & $2.14$                            & $5\pm1$     \\
\hline
$100\,\mathrm{nm}$ & 3 layers  &  1 (oxide)                  &   $3.9\pm0.5$  & $(3.52\pm0.06)-(0.8\pm0.2)i$     & $3.6\pm0.4$  \\
                   &           &  2 ($\mathrm{^{10}B_4C}$)    &   $115\pm2$    & $(2.11\pm0.06)-(1.18\pm0.05)i$   & $4\pm1$   \\
                   &           &  3 (substrate)              &   -            & $2.14$                         & $5\pm1$   \\
\hline
 $1\,\mathrm{\mu m}$ & 1 layer  &  1 ($\mathrm{^{10}B_4C}$)   &   -            & $(2.48\pm0.05)-(1.01\pm0.04)i$  & $3.1\pm0.3$  \\
\hline
 $1\,\mathrm{\mu m}$ & 2 layers  &  1 (oxide)                &   $3.5\pm0.5$  & $(3.15\pm0.06)-(0.5\pm0.1)i$    & $4.6\pm0.4$ \\
                     &           &  2 ($\mathrm{^{10}B_4C}$)  &   -            & $(2.08\pm0.05)-(1.10\pm0.04)i$  & $2.1\pm0.3$   \\
\hline \hline
\end{tabular}
\end{table}
\section{Reflection of neutrons by converters at grazing angle used in neutron detection}
\begin{figure}[!ht]
\centering
\includegraphics[width=10cm,keepaspectratio]{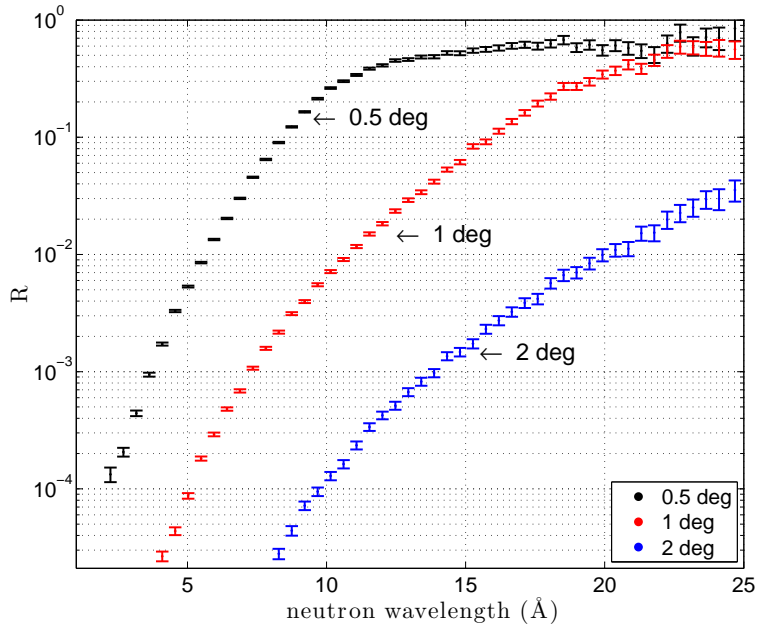}
\caption{\footnotesize Measured reflectivity of $1\,\mu m$ $^{10}B_4C$ deposited on Si as a function of the neutron wavelength ($\lambda$) for 3 different angles.}\label{ertdb87}
\end{figure}

The efficiency of a thermal neutron detector exploiting a solid neutron converter increases rapidly as the angle between the incoming neutrons and the converter decreases below 10 degrees. Detailed analytical calculations are given in~\cite{frath}. It has also been demonstrated by experimental results~\cite{framb,kleinjalousie,kleinjalousie2,nowak}. As already mentioned the neutron reflection must be avoided in neutron detection because it limits the maximum efficiency that can be attained. Moreover, the detector requirements for low sensitivity to background~\cite{esstdr,kir4} are becoming more and more strict~\cite{estia1,estia2,freia}. Background events can arise from $\gamma$-ray detection or background neutrons that can give rise to misaddressed events. The $\gamma$-ray sensitivity must be kept very low with respect to neutron efficiency, typically $10^{-6}$~\cite{gammas}. The same order of magnitude must be kept for the neutron background considering that the neutron detection efficiency is much larger than that of $\gamma$-rays. The neutrons that are reflected by the converter in a detector can strongly contribute to the background. Therefore, the neutron reflection must be taken into account in the detector concept even if it is very low. 
\\We characterized the $\mathrm{^{10}B_4C}$ layers deposited on Si; the converter roughness on this substrate is a few nm due to low surface roughness of the substrate. Figure~\ref{ertdb87} shows the reflectivity curve for the $1\,\mathrm{\mu m}$ sample on Si as a function of the neutron wavelength (ToF measurement) for three different angles (0.5, 1 and 2 degrees). Note that for wavelengths larger than $20$\AA, if we use a converter inclined at 1 degree (red curve) about $30\%$ or more of the neutrons are reflected, thus not converted. Already at 2 degrees, the reflection is negligible for most potential applications. 
\\For the $1\,\mathrm{\mu m}$-thick sample deposited on Al no specular reflection was observed or the reflected neutron intensity was below the background of the instrument ($\sim10^{-6}$) at any value of q. The specular reflection is attenuated by the surface roughness. In order to diminish the reflection effect in a detector, it is sufficient to have a rough surface, as for Al where it is a few tens of nm. This can be of importance for detectors based on micro-strips and solid converters. Operated at small angle, the absorber deposition on glass could not have a large enough roughness to avoid significant reflection. It has to be pointed out that an excessive roughness will also degrade the efficiency at small angles~\cite{framb,irinamac}. When the roughness becomes comparable to the neutron capture fragments path lengths in the converter ($\sim1\,\mu m$ for $\mathrm{^{10}B_4C}$) the surface can not be considered flat anymore. The flatness is essential for the neutron to traverse a large thickness of $^{10}B$, and the conversion fragments to be close to the surface to be able to escape. If the roughness starts to be comparable to the conversion fragments ranges, this assumption is not valid anymore and it results in a drop in the expected efficiency.
\section{Conclusions}
A new technique to measure and exploit the neutron reflectivity along with the neutron-induced fluorescence for layers of strongly neutron absorbing materials has been described. The theoretical model has been developed for single and multiple layers and it has been understood in the light of the measurements performed on $\mathrm{^{10}B_4C}$ films deposited on both Si and Al. The understanding of the partially oxidized surface layer is crucial to obtain a fit with reasonable physical constraints. Film thickness, scattering length densities and surface roughnessess can be determined from the fit. We extended the neutron-induced fluorescence technique to measure and fit the absorption over a wide q-range not only limited below the critical edge. Neutron reflectometry measured together with fluorescence is a powerful non-destructive tool to directly obtain the number density of the absorbing isotope, $\mathrm{^{10}B}$ in our experiment. 
\\We characterized the $\mathrm{^{10}B_4C}$ layers in order to understand the amount of reflection that must be minimized for application in neutron detection. Surface roughness helps to attenuate the reflection as it was observed on the film deposited on Al. At a very grazing angle ($\approx1\,$degree) the reflection does not only reduce the maximum detection efficiency that can be attained but can also generate a source of background that must be taken in into account in the detector concept.  
\\Finally, we note that the methods and theory developed open up the ability of using neutron reflection as a diagnostic technique on highly absorbing films.
\bigskip
\textbf{\large \\ Acknowledgements}
\medskip
\\ This work was carried out as a part of the collaboration between the ILL, ESS, and Link\"{o}ping University on developing $^{10}B$ thin film neutron detectors, within the context of the International Collaboration on the development of Neutron Detectors (www.icnd.org).
\\ The work has been also partially supported by the CRISP project (European Commission 7th Framework Programme Grant Agreement 283745).
\\ Beamtime on SuperAdam and A.J.C. Dennison were funded by the Swedish Research Council VR 2009-6232 (ILL research proposal 58782 - Reflectivity of 10B4C neutron converter films for novel neutron detectors).
\\ S. Schmidt acknowledges the support by the Carl Tryggers Foundation for Scientific Research.
\\ J. Birch is grateful for the support from Knut and Alice Wallenberg's Foundation through the project grant: "Isotopic Control for Ultimate Material Properties".
\\ The author wish to thank M. Jentschel for the support with the HPGe detector and B. Toperverg for valuable discussions.


\begin{thebibliography}{alp}
\bibitem{cubitt3}  Roy Pike and Pierre Sabatier, \emph{Scattering - Scattering and inverse scattering in Pure and Applied Science} (Academic Press, London, 2002), Vol. 2, p. 1198-1208.
\bibitem{fermizinn} E. Fermi and W. H. Zinn, \emph{Reflection of neutrons on mirrors}. Los Alamos National Laboratory, U.S. Atomic Energy Commission, 1946.
\bibitem{padlo} V. Aksenov \emph{et al.}, Physica B \textbf{276-278}, 946-947 (2000).
\bibitem{nistbr} H. Zhang \emph{et al.}, Phys. Rev. Lett. \textbf{72}, 3044-3047 (1994). 
\bibitem{manu2} E. Schneck \emph{et al.}, Langmuir \textbf{29}, 4084-4091 (2013).
\bibitem{esstdr} S. Peggs \emph{et al.}, \emph{ESS Technical Design Report}, ESS-doc-274, 2013.
\bibitem{kir4} O. Kirstein \emph{et al.}, arXiv:1411.6194, (2014).
\bibitem{ChoScience}A. Cho, Science \textbf{326}-5954, 778-779 (2009).
\bibitem{KarlNN} K. Zeitelhack, Neutron News \textbf{23}-10, 10-13 (2010).
\bibitem{blue} Albert-Jose Dianoux and Gerry Lander, \emph{Neutron Data Booklet, Second Edition} (2013).
\bibitem{buff3} J.C. Buffet \emph{et al.}, in Conference record of Nuclear Science Symposium and Medical Imaging Conference (NSS/MIC) Anaheim, IEEE Trans. Nucl. Sci. 171-175 (2012).
\bibitem{framb} F. Piscitelli \emph{et al.}, J. Instrum. \textbf{9}, P03007 (2014).
\bibitem{fratesi}  F. Piscitelli,  Ph.D. thesis, Institut Laue-Langevin and University of Perugia, 2014.
\bibitem{mbmg} F. Piscitelli, Eur. Phys. J. Plus \textbf{130}:27 (2015).
\bibitem{kleinjalousie} M. Henske \emph{et al.}, Nucl. Instrum. Meth. A \textbf{686}, 151-155 (2012).
\bibitem{kleinjalousie2} G. Modzel \emph{et al.}, Nucl. Instrum. Meth. A \textbf{743}, 90-95 (2014).
\bibitem{nowak} G. Nowak \emph{et al.}, J. Appl. Phys. \textbf{117}, 034901 (2015).
\bibitem{vanvuure} T. L. van Vuure \emph{et al.}, IEEE Trans. Nucl. Sci. \textbf{57}, 323-327 (2010).
\bibitem{gor} G. Croci \emph{et al.}, Europhys. Lett. \textbf{107}, 12001 (2014).
\bibitem{carina} C. H\"{o}glund \emph{et al.}, J. Appl. Phys. \textbf{111}, 10490-8 (2012).
\bibitem{sears}  V. F. Sears, Neutron News \textbf{3}:3, 29-37 (1992).
\bibitem{hayter} J. B. Hayter. and H. A. Mook, J. Appl. Crystallogr. \textbf{22}, 35-41 (1989).
\bibitem{schiff} L. I. Schiff, \emph{Quantum Mechanics} - third edition. McGraw-Hill, Inc. (1955).
\bibitem{parratt} L. G. Parratt, Phys. Rev. \textbf{95}, 359 (1954).
\bibitem{carinatesi} C. H\"{o}glund, Ph.D. thesis, Link\"{o}ping University - Institut of Technology, 2010.
\bibitem{carrad} C. H\"{o}glund \emph{et al.}, Radiat. Phys. Chem. \textbf{113}, 14-19 (2015).
\bibitem{mg} J. Birch \emph{et al.}, IEEE Trans. Nucl. Sci. \textbf{PP}:9, 1-8 (2013).
\bibitem{cubittD17} R. Cubitt \emph{et al.}, Appl. Phys. A \textbf{74}, s329-s331 (2002).
\bibitem{superadam} A. Devishvili \emph{et al.}, Rev. Sci. Instrum. \textbf{84}, 025112 (2013).
\bibitem{frath} F. Piscitelli \emph{et al.}, J. Instrum. \textbf{8}, P04020 (2013).
\bibitem{estia1} J. Stahn \emph{et al.}, Eur. Phys. J. Appl. Phys. \textbf{58} 11001 (2012).
\bibitem{estia2} J. Stahn, \emph{ESTIA: A Truly Focusing Reflectometer}. ESS instrument proposal (2014).
\bibitem{freia} H. Wacklin, \emph{FREIA: Reflectometer concept for fast kinetics at ESS}. ESS instrument proposal (2014).
\bibitem{gammas} A. Khaplanov \emph{et al.}, J. Instrum. \textbf{8}, P10025 (2013).
\bibitem{irinamac} I. Stefanescu \emph{et al.}, Nucl. Instrum. Meth. A \textbf{727}, 109-125 (2013).
\end{thebibliography}
\end{document}